\documentclass[aps,pra,twocolumn,showpacs,superscriptaddress,preprintnumbers,amsmath,amssymb,floatfix,fnpara]{revtex4}
\usepackage{graphicx}                                                                           % Include figure files
\usepackage{dcolumn}                                                            % Align table columns on decimal point
\usepackage{bm}                                                                                            % bold math
\usepackage{color}                                                                                      % colored text
\newcommand{\bc}{\begin{center}}
\newcommand{\ec}{\end{center}}
\newcommand{\ba}{\begin{array}}
\newcommand{\ea}{\end{array}}
\newcommand{\fnm}{\footnotemark}
\newcommand{\fnt}{\footnotetext}

\begin{document}
%===============================%
%===%        AUTHORS        %===%
%===============================%
\author{M.~F. Hasoglu}
%\altaffiliation{Present address, Gazikent University}
\affiliation{Georgia State University, Atlanta, GA 30303}
\affiliation{Gazikent University, 27100 Sahinbey, Gaziantep, Turkey }
\author{T.~W. Gorczyca}
\affiliation{Western Michigan University, Kalamazoo, MI 49008}
\author{M.~A. Bautista}
\affiliation{Western Michigan University, Kalamazoo, MI 49008}
\author{Z.~Felfli}
\affiliation{Center for Theoretical Studies of
        Physical Systems, Clark Atlanta University, Atlanta, GA 30314}
\author{S.~T. Manson}
\affiliation{Georgia State University, Atlanta, GA 30303}
%===============================%
%===%         TITLE         %===%
%===============================%
\title{Radiation Damping in the Photoionization of Fe$^{14+}$}
%===============================%
%===%         DATUM         %===%
%===============================%
\date{\today}
%===============================%
%===%        ABSTRACT       %===%
%===============================%
\begin{abstract}

A theoretical investigation of photoabsorption and photoionization of Fe$^{14+}$ extending beyond an earlier frame transformation R-matrix implementation is performed using a fully-correlated, Breit-Pauli R-matrix formulation including both fine-structure splitting of strongly-bound resonances and radiation damping. The radiation damping of $2p\rightarrow nd$ resonances gives rise to a resonant photoionization cross section that is significantly lower than the total photoabsorption cross section. Furthermore, the radiation-damped photoionization cross section is found to be in good agreement with recent experimental results once a global shift in energy of $\approx -3.5$ eV is applied. These findings have important implications. Firstly, the presently available synchrotron experimental data are applicable only to photoionization processes and not to photoabsorption; the latter is required in  opacity calculations. Secondly, our computed cross section, for which the L-shell ionization threshold is aligned with the NIST value, shows a series of $2p \rightarrow nd$ Rydberg resonances that are uniformly 3-4 eV higher in energy than the corresponding experimental profiles, indicating that the L-shell threshold energy values currently recommended by NIST are likely in error.

\end{abstract}
\maketitle

Most of what we know about our universe is obtained through spectroscopy.
 We study either the emission of hot plasma sources or the absorption of  intervening gas between us and a bright object.
  For example, it is possible to probe the multiphase interstellar medium (ISM) through the observation of atomic absorption lines and edges in the spectra of background sources, and the absorption properties of the plasma depend on the {\it photoabsorption} cross sections of the species present \cite{kallman}.
To interpret emission spectra, on the other hand, one first needs to determine the ionization structure of the gas, which, for photoionized plasmas, depends on the {\it photoionization} cross sections of the chemical elements present.
Unlike spectra in the visual region dominated by valence electron processes in low ionization species, extreme UV and X-ray spectra show inner L- and K-shell processes for which {\it photoionization} and {\it photoabsorption} are significantly different.
The difference between these two processes arises from the fact that absorption of a photon by the atom to an autoionizing state (above the threshold) does not necessarily lead to electron emission.

For the present case of photons incident on Fe$^{14+}$ in the vicinity of the L-edge,
the difference between photoabsorption and photoionization can be understood by
considering the prominent $2p\rightarrow nd$ absorption resonances.
Following absorption of a photon,
\begin{eqnarray}      \label{eq1}
h\nu + 2p^63s^2 & \longrightarrow & 2p^53s^2nd  \ ,
\end{eqnarray}
an intermediate resonance state can either
decay via autoionization, or Auger decay, leading to ionization of Fe$^{14+}$ into the Fe$^{15+}$ ion plus a free electron, or it can radiatively stabilize, leading to no charge change of the Fe$^{14+}$ ion:
\begin{eqnarray}      \label{eq2}
 2p^53s^2nd &  \longrightarrow &
        2p^63s+ e^-  \   (A_a^{part}\sim n^{-3}) \label{aapart}\\%\nonumber\\
&  \longrightarrow &      2p^6nd+ e^-  \ (A_a^{spect}\ {\rm ind.\ of\ }n) \label{aaspect}\\%\nonumber\\
&  \longrightarrow & 2p^63s^2 + h\nu^\prime \  (A_r^{part}\sim n^{-3}) \label{arpart}\\%\nonumber\\
&  \longrightarrow & 2p^63snd + h\nu^\prime \  (A_r^{spect}\ {\rm ind.\ of\ }n)\label{arspect}
\end{eqnarray}
Thus, the alternative decay pathways of radiative stabilization redirect some of the initial photoabsorption amplitude, giving a reduced, or {\em damped} \cite{burke},
photoionization cross section relative to the photoabsorption cross section.

The extent of this so-called radiation damping effect, for an isolated resonance, is essentially given by the autoionization branching ratio
\begin{eqnarray}
\label{branch}
\frac{\sigma_{PI}}{\sigma_{PA}} & \approx & \frac{A_a^{tot}}{A_a^{tot}+A_r^{tot}}\ ,
\end{eqnarray}
where $\sigma_{PI}$ and $\sigma_{PA}$ are the photoionization and photoabsorption cross sections, respectively. The total
autoionization and radiative rates,
$A_a^{tot}$ and $A_r^{tot}$, are each the sum of
contributions for which the $nd$ Rydberg electron {\em participates}, with partial rates scaling as $n^{-3}$, and for which the Rydberg electron is a {\em spectator}, with partial rates that are independent of $n$.
Thus, the spectator Auger and radiative decay branches (Eqs. \ref{aaspect} and \ref{arspect}) dominate as $n\rightarrow \infty$.  On the other hand, due to angular momentum, or geometric, and radial, or
dynamic, considerations of the $3d$ orbital compared to the $3s$ orbital,
the $2p^53s^23d$ resonance decays predominantly via participator pathways, as
given in Eqs.
\ref{aapart} and \ref{arpart}.

To date, essentially all of the calculated photon-atom data for multiply-charged ions has
been produced by theoretical photoionization calculations
without considering the alternate radiative decay pathway in Eqs.~\ref{arpart} and \ref{arspect} (see, for example, Refs.~\cite{opacity1,opacity2}).
These undamped cross sections are equivalent to the photoabsorption cross sections regarding absorption strength (although the resonance width is underestimated),
and thus it had always been assumed that $\sigma_{PI}=\sigma_{PA}$.
For Fe$^{14+}$, perhaps the most
definitive calculations to date are the R-matrix photoionization calculations in both LS coupling \cite{bautista} and using an LS-coupled to a JK-coupled frame transformation (LSJKFT) method, along with resonance broadening in order to correctly characterize the resonances \cite{gorczyca}.
In that latter study (hereafter referred to as Paper I), it was demonstrated that the LSJKFT method was capable of reproducing all fine-structure splitting effects of a full Breit-Pauli R-matrix (BPRM) calculation except for the lowest $2s^22p^53s^23d$ and $2s2p^63s^23p$ resonances.
This is because in the LSJKFT method, an extremely efficient computational approach compared to the BPRM method, the multi-channel quantum defect
(MQDT) equations for the outer-region solutions are modified, thereby incorporating
fine-structure splitting only for those resonances that reside outside the R-matrix "box"; those that reside within the R-matrix region (i.e., all states
described by orbitals only up to $n=3$, such as the $2s^22p^53s^23d$ resonances)
did not include fine-structure splitting effects.

The agreement between LSJKFT and BPRM methods (except for the strongly-bound resonances) was first demonstrated only for a minimal configuration-interaction (CI) case, since the large-scale CI calculation which would have been required for a more converged calculation and which was performed only within an LSJKFT approach,
would have been prohibitively large to undertake within a full BPRM approach.  Interestingly, it was found
that, in terms of general qualitative features, the large-CI LSJKFT cross section,
convoluted with a sufficiently broad energy distribution, was essentially the same as that from a minimal-CI LS calculation.  Thus,  it is necessary to include higher-order CI and fine-structure effects only if a detailed resonance
description is desired.  It should be noted that all of those earlier calculations were for the total photoabsorption cross section, and did not consider radiation damping effects.  Furthermore,
those earlier cross sections were all preconvoluted \cite{francon} with a constant Lorentzian width of 0.1 Ryd \cite{typoI}, so a detailed
investigation of the resonance profiles was not performed.

More recently, an electron-beam ion trap (EBIT) was used at a synchrotron facility to detect the final production of Fe$^{15+}$ ions \cite{fe14exp}, thereby measuring the {\em photoionization} cross section at an energy resolution as low as 150 meV (see also Ref.~\cite{ebit2}).
Thus, it is now desirable to compute {\em radiation-damped} photoionization cross sections, since only total photoabsorption cross sections were reported in Paper I.  Furthermore, to compare directly with the finer resolution of the experimental data, it is necessary to consider higher-order CI effects
and  fine-structure splitting for all resonances, including the lowest $2s^22p^53s^23d$ resonances for which a frame transformation approach is inapplicable.
The present study improves upon the theoretical work of Paper I
and other earlier calculations \cite{gu,fe14exp} by using the same large-scale CI description of Paper I within a full BPRM calculation, which treats fine-structure splitting correctly for even the lowest resonances, and by including radiation damping effects \cite{frandamping}.  The
BPRM calculations are performed using the standard
UCL/Belfast R-matrix codes \cite{burke} with additional modifications for incorporating radiation damping \cite{frandamping}, as described in earlier electron-impact excitation \cite{ti20} and dielectronic recombination
\cite{ar15} studies.  A brief description of the present methodology is given below, followed by a detailed comparison between these new results and the recent experimental results.

The atomic structure used is essentially the same as that of Paper I. %\cite{typoII}.
The Fe$^{15+}$ target states
are described by an orbital basis set that is generated from a
Hartree-Fock \cite{ffmchf} calculation on the $1s^22s^22p^63s$ ground state configuration, with additional $3p$ and $3d$ orbitals generated from frozen-core
Hartree-Fock calculations on the $1s^22s^22p^63p$ and $1s^22s^22p^63d$
excited configurations, respectively.
All target states of the form $2s^22p^63l$, $2s^22p^53s3l$, and
$2s2p^63s3l$ are used in the close-coupling expansion and are described by a larger CI basis consisting of all
configurations consistent with single and double promotions
out of those three base configurations.

The corresponding ($N+1$)-electron bound, continuum, and resonance states of Fe$^{14+}$ are described by a basis consisting of single, double, and triple
promotions out of the $2s^22p^63s^2$ base configuration.
Lastly, unlike in Paper I, corrections to the $N$-electron Hamiltonian are applied in order to realign the target energies with the recommended NIST \cite{nist} values
(see Table I), and the binding energy of the Fe$^{14+}$ ground state, relative to the Fe$^{15+}$ state, is also aligned with the NIST value.
Thus, of particular importance,
the present theoretical photon energy thresholds which we report
are aligned with the corresponding NIST values.

The important details regarding radiation damping are that all
$2s^22p^63s^2$ and $2s^22p^63s3d$ ($J=0,1,2$) bound states are
included as final type II (inner-region) radiative decay states,
resulting in a complex inner-region R-matrix \cite{frandamping,ti20,ar15}.
This accounts for all the participator radiative decay pathways in Eq.~\ref{arpart} and  spectator radiative decay of the $2p^53s^23d$ resonance.
The spectator radiative decay of all higher $nd$ resonances in Eq.~\ref{arspect},
on the other hand, is accounted for by a modification to the MQDT
equations in the outer region \cite{frandamping,ti20,ar15}, where the
 type I ($3s\rightarrow 2p$ core) decay width used in the $E\rightarrow E-i\Gamma_r/2$ modification
 is  $\Gamma_r= 0.45$ meV.

The target states included explicitly in our R-matrix calculation only account for the participator Auger decay of Eq.~\ref{aapart} and the $n=3$ spectator Auger decay of Eq.~\ref{aaspect}
to the so-called {\em main line} photoionization continua.
The {\em satellite} photoionization continua of Eq.~\ref{aaspect} for $n>3$
are instead included in our formulation via an optical potential MQDT modification procedure  \cite{ar},
similar to that done for the spectator radiative decay.
This presents a difficulty in extracting a total theoretical photoionization cross section, since the above described methodology for including
radiation damping {\em and} $n>3$ spectator Auger decay yields only total photoabsorption and main-line photoionization cross sections \cite{ar}.
The difference between the two includes both satellite photoionization
and radiative damping amplitudes. As $n\rightarrow\infty$, the branching of the two can be extracted by using the spectator rates in Eq.~\ref{branch}, whereas for low-$n$, the difference is due purely to radiation damping.  For intermediate $n$, however, the branching requires a deeper investigation beyond the scope of the present work.

In Fig.~1, the present BPRM results are compared to the earlier results of Paper I and to recent experiment \cite{fe14exp}.
In Fig.~1a, our new BPRM photoabsorption results, convoluted with a Lorentzian width of 0.1 Ryd, are found to be qualitatively similar to
the earlier LSJKFT results, but differ quantitatively in three
minor respects.   First, there is a
uniform energy difference between the two since the earlier results were computed using theoretical thresholds (NIST values were not available for
the $2p^53s^2$ autoionizing states at that time) whereas our present results have been aligned to the NIST thresholds.
Second, the LSJKFT results were preconvoluted \cite{francon} with an 0.1 Ryd Lorentzian \cite{typoI} only above 931.55 eV, whereas the present BPRM results
have preconvoluted even the lowest resonances.
Third, and most importantly, the LSJKFT results do not account for
fine-structure splitting of the lowest $3d$ resonances, as noted earlier.
The BPRM and LSJKFT cross sections are also compared to each other in Fig.~1b on a linear scale,
convoluted with a Voigt profile   (Lorentzian and Gaussian widths $\Gamma_L=2.11$ eV and $\Gamma_G=4.0$ eV, respectively). It is seen that,
except for the slight global energy shift and the BPRM fine-structure splitting of the
$2s^22p^53s^23d$ resonances,
the two cross sections are nearly identical.  Thus, the only improvement over the LSJKFT method by using instead a BPRM method is the inclusion
of fine-structure splitting of the lowest resonances, as had been
noted in Paper I.

\begin{figure}[thb]
\includegraphics[scale=0.65,angle=0.]{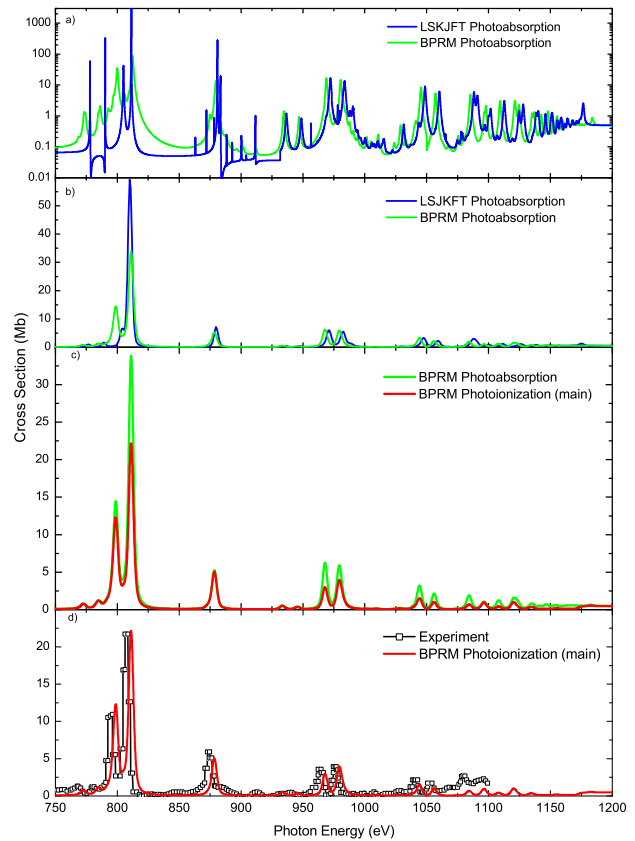}\vspace*{-1mm}
\caption{\label{fig1}
(Color online) Fe$^{14+}$ photoabsorption and main-line photoionization cross sections in the vicinity of the L-edge.
a) Earlier LSJKFT photoabsorption results \cite{gorczyca} (blue line) and
present BPRM photoabsorption results (green line), using a  Lorentzian width of 0.1 Ryd.
b) Same as a) using a Voigt profile ($\Gamma_L=2.11$ eV, $\Gamma_G=4.0$ eV).
c)
Present BPRM photoabsorption results (green line) and main-line photoionization results (red line) using the same Voigt profile.
b) Present main-line photoionization results (red line)
and experimental results \cite{fe14exp} (black data points).
 }
\end{figure}

Using the same resolution, we compare our  total photoabsorption cross section
to the computed main-line, damped photoionization cross section
in Fig.~1c.  As noted above, the  main-line photoionization cross section for the $n=3$ resonances below 950 eV photon energies, and, for the most part, the
$n=4$ resonances, accounts for the total photoionization cross section,
and thus there is appreciable $3d\rightarrow 2p$ radiation damping of the lowest $2p^53s^2nd$ resonances;
the $2s2p^63s^23p$ resonance at 880 eV, on the other hand,  is not damped significantly due to the much smaller $3p\rightarrow 2s$ radiative rate.

We compare the main-line photoionization cross sections to the experimental results (arbitrarily scaled) in Fig.~1d. Except for
an approximately constant energy difference, good agreement between the two results is found in resonance profiles up to about 1050 eV.
It is interesting to note that the relative heights of the
$2p^53s^2(^2P_{3/2})3d$ and $2p^53s^2(^2P_{1/2})3d$ resonances are the same in both the BPRM photoionization  and experimental cross sections, whereas the BPRM photoabsorption cross section indicates a much different height ratio.

The comparison of relative $2p^53s^23d$ resonance heights would seem to validate experimentally the extent of radiation damping.
Nevertheless, we can get a more quantitative assessment by examining
{\em absolute} resonance strengths.
By fitting our photoionization and photoabsorption cross sections, in the vicinity of the $2p^53s^2(^2P_{1/2})3d$ resonance, to an
energy-normalized Lorentzian multiplied by a strength factor, we obtained integrated resonance strengths of 130 Mb-eV and 200 Mb-eV, respectively.
Table I indicates that our photoionization resonance strength is indeed
in agreement with the experimental value of 110$\pm$60 Mb-eV
whereas the photoabsorption resonance strength is outside of the experimental uncertainty, providing further validation of the extent of radiation damping.

\begingroup
\squeezetable
\begin{table}[bthp]
    \caption{\label{table1}
$2p^53s^2(^2P_{1/2})3d$ Resonance Strengths (Mb-eV).}
    \begin{ruledtabular} \begin{tabular}{ll}
    Experiment \cite{fe14exp} & 110 $\pm$ 60 \\
    R-matrix Photoionization & 130 \\
    R-matrix Photoabsorption& 200
    \end{tabular}
    \end{ruledtabular}
\end{table}
\endgroup

\begingroup
\squeezetable
\begin{table}[bthp]
    \caption{\label{table2}
Fe$^{14+}$ Resonance Energies and Fe$^{15+}$ Thresholds (eV).}
    \begin{ruledtabular} \begin{tabular}{lrrrr}
Level &     Exp.~\cite{fe14exp} & Present & NIST~\cite{nist} & Exp.~\cite{graf} \\
%\fnm[1]
    \colrule
Fe$^{14+}(2p^53s^2(^2P_{3/2})3d)$ & 794.7 &  798.77 &   &   \\
Fe$^{14+}(2p^53s^2(^2P_{3/2})4d)$ & 964.3 &  967.81 &   &   \\
Fe$^{14+}(2p^53s^2(^2P_{3/2})5d)$ & 1040.9   & 1044.22  &   &   \\
\ \ \ \ \vdots  & & & \\
Fe$^{15+}(2p^53s^2(^2P_{3/2}))$ &    & 1172.76\fnm[1]  &  1172.76   & 1170.88  \\
    \colrule
Fe$^{14+}(2p^53s^2(^2P_{1/2})3d)$ & 807.1   &  811.13 &   &   \\
Fe$^{14+}(2p^53s^2(^2P_{1/2})4d)$ & 976.0   &  979.25 &   &   \\
Fe$^{14+}(2p^53s^2(^2P_{1/2})5d)$ & 1053.5   &   1055.9 &   \\
\ \ \ \ \vdots  & & & \\
Fe$^{15+}(2p^53s^2(^2P_{1/2}))$ &    & 1185.16\fnm[1]  & 1185.16  &   \\
    \end{tabular}
    \end{ruledtabular}
\fnt[1]{Theoretical thresholds aligned to NIST values \cite{nist}.}
%\fnt[2]{Present R-matrix calculations with threshold aligned to NIST value.}
%\fnt[3]{\cite{nist}.}
%\fnt[4]{\cite{graf}.}
\end{table}
\endgroup

The comparison in Fig.~1d also indicates
that the BPRM resonance energies are uniformly higher than the corresponding experimental values,
even though the BPRM cross sections have been  aligned to the
 $n\rightarrow\infty$ thresholds of the $2p^53s^2(^2P_{3/2})nd$
and $2p^53s^2(^2P_{1/2})nd$ Rydberg series given by NIST as 1172.77 and 1185.17 eV, respectively.  Table II lists the resonance energies, indicating that the resonance energy difference does not approach zero with the expected $1/n^3$ behavior, assuming the same $n\rightarrow\infty$ Rydberg limits.
This seems to suggest that our threshold values used - those recommended by NIST \cite{burkhalter,yuri} - are too high.
Indeed, a more recent experiment \cite{graf} and subsequent theoretical analysis \cite{beiersdorfer} reports a lower $2p^53s^2(^2P_{3/2})$ threshold of 1170.88 eV,
which would account for about 2 eV of the difference in resonance energy positions.  Although newer measurements for the $2p^53s^2(^2P_{1/2})$
threshold energy are not available, it is likely that this NIST value is also in error.

In conclusion, we have reported new R-matrix calculations that improved on our earlier work of Paper I by including fine-structure effects for the lowest resonance and radiation damping effects for all resonances.
We find that radiation damping is significant for the $2p^53s^2(^2P_{1/2})3d$ resonance, in particular.  Our radiation-damped photoionization resonance strengths show good agreement with the experimental results, but there is a nearly uniform difference in energy positions between the present theoretical and the experimental \cite{fe14exp} resonance positions.  Since the present calculations have been aligned to the NIST
experimental threshold value \cite{burkhalter}, there is an apparent
inconsistency between theoretical and experimental resonance positions which is not fully resolved from a newer threshold measurement \cite{graf}.

We thank  M. C. Simon and J. R. Crespo L\'{o}pez-Urrutia for fruitful discussions and for providing the experimental results. This work was supported in part by a NASA~APRA grant, and in part by DOE, Office of Chemical Sciences.

%===============================%
%===%      BIBLIOGRAPHY     %===%
%===============================%
\bibliographystyle{apsrev.bst}
%\bibliography{fe14}

\begin{thebibliography}{21}
\bibitem[{typ()}]{typoI}
\bibinfo{note}{In Paper I, it was erroneously stated that the Lorentzian
  preconvolution width used was 1.0 Ryd whereas the actual width used was 0.1
  Ryd.}
\end{thebibliography}

\end{document}